\documentclass[12pt, preprint]{aastex}

\newcommand{\mum}{\ifmmode{\mu m}\else{$\mu$m}\fi}
\newcommand{\iso}{{\em ISO}}
\newcommand{\iras}{{\em IRAS}}

\begin{document}
\title{Artifacts at 4.5 and 8.0 \mum\ in Short Wavelength Spectra from the
{\em Infrared Space Observatory}\footnote{Based on observations with
the {\em Infrared Space Observatory (ISO)}, an European Space Agency
(ESA) project with instruments funded by ESA Member States (especially
the Principal Investigator countries: France, Germany, the Netherlands,
and the United Kingdom) and with the participation of the Institute
of Space and Astronautical Science and the National Aeronautics and
Space Administration (NASA).}
}

\author{Stephan D. Price\altaffilmark{2}, G. C. 
Sloan\altaffilmark{3,~4},
Kathleen E. Kraemer\altaffilmark{2,~5}}

\altaffiltext{2}{Air Force Research Laboratory, Space Vehicles 
Directorate,
           29 Randolph Rd., Hanscom AFB, MA 01731-3010; 
	steve.price@hanscom.af.mil, kathleen.kraemer@hanscom.af.mil}
\altaffiltext{3}{Institute for Scientific Research, Boston College,
           140 Commonwealth Ave., Chestnut Hill, MA 02467-3862;
	sloan@ssa1.arc.nasa.gov}
\altaffiltext{4}{Infrared Spectrograph Science Center, Cornell 
University,
           Ithaca, NY 14853-6801}
\altaffiltext{5}{Institute for Astrophysical Research, Boston University,
           Boston, MA 02215}

\begin{abstract}
Spectra from the Short Wavelength Spectrometer 
(SWS) on \iso\ exhibit artifacts at 4.5 and 8~$\mu$m.  These 
artifacts appear in spectra from a recent data release, OLP 10.0, 
as spurious broad emission features in the spectra 
of stars earlier than $\sim$F0, such as $\alpha$ CMa.  Comparison of 
absolutely calibrated spectra of standard stars to corresponding 
spectra from the SWS reveals that these artifacts result from an 
underestimation of the strength of the CO and SiO molecular bands 
in the spectra of sources used as calibrators by the SWS. Although
OLP 10.0 was intended to be the final data release, these
findings have led to an additional release addressing this issue, OLP 10.1, 
which corrects the artifacts.

\end{abstract}


\section{Introduction \label{sec.intro}}

The Short Wavelength Spectrometer (SWS) on the {\em Infrared Space
Observatory (ISO)} obtained approximately 1250 spectra covering the 
full 2.4--45~$\mu$m wavelength range at moderate resolution.  We are 
engaged in an ongoing project to classify these spectra \citep{ksp01}, 
to reprocess them, and to present them in a publically available 
database on-line (Sloan et al. 2002, in preparation).  SWS data are 
currently available in a partially processed form called Auto-Analysis 
Results (AARs).  The AARs have a fairly complex format \citep{lee01} 
that requires further processing before the data can be used for 
scientific analysis.  Our processing  method reduces these data to a 
single continuous 2.4 to 45~$\mu$m spectrum.

In assessing the quality of our new reprocessing algorithm, 
we compared the results for several infrared standards observed with 
the SWS to absolutely calibrated spectra from Cohen et al. (1992a, 
1992b, 1995, 1996a, 1996b, 2001 (in preparation)).  These standards 
are based on synthetic spectra of the A0 dwarf $\alpha$ Lyr and the 
A1 dwarf $\alpha$ CMa, which serve as the reference standards in the
system.  Cohen et al. (1992a) describe the details of their method that
used high-quality ground-based and airborne photometry to normalize the
synthetic spectra to measured astronomical fluxes.  The synthetic
spectra are based on the models of \cite{kur79}, with updated
opacities and metallicities \citep{ck94}.

Secondary standards are added to the system by observing their spectra 
in conjunction with that of the primaries so that atmospheric, 
telescopic, and instrumental transients can be removed.  The spectra
for the secondary standards are then obtained by dividing the observed
secondary by the observed reference standard and multiplying by the
assumed spectrum (the model) for the reference, degraded to match the
spectral resolution of the instrument used:

\begin{eqnarray}
  S_{b,final}(\lambda) =
  \frac{S_{b,obs}(\lambda)}{S_{a,obs}(\lambda)}
 S_{a,assumed}(\lambda),
\label{eqn.1}
\end{eqnarray}

\noindent where the subscript $a$ refers to the reference standard,
and the subscript $b$ refers to the secondary.  This is the standard
method used by spectroscopists at ground-based telescopes to 
calibrate a program source by ratioing its spectrum to that of a 
standard star, preferably secured at an airmass matching that of the
target.  In effect, Equation \ref{eqn.1} transfers the 
quality of the synthetic A star model from the reference to the new 
standard, whatever its spectral type.

Cohen et al. (1992b, 1995, 1996a, 1996b, 2001) have applied this
method to create absolutely calibrated composite spectra for 13
infrared standards, using spectra from ground-based telescopes, the
Kuiper Airborne Observatory (KAO) and the {\em Infrared Astronomical
Satellite (IRAS)} \citep[further details of the process can be found in 
][]{cwb92a, cww92b}. \cite{cwb96a} added $\alpha^{1}$ Cen (G2 V) as an
alternative reference standard for the southern hemisphere to give a
total of 3 reference standards as well as the 13 secondaries.  Most
of the secondary standards are giants with spectral classes later
than K0 and were chosen for their intrinsic brightness; sources later 
than approximately M3 are avoided due to the possibility of  their 
variability \citep{eg97}.

The preferred infrared reference standard is Sirius ($\alpha$ CMa),
due to its brightness and its dust-free spectrum beyond 20~$\mu$m.
Figure 1 compares our derived SWS spectrum of this source to the
Kurucz model presented by Cohen et al. (1992a).  Deviations between
the two occur in the vicinity of 4.5~$\mu$m and again at 8.0~$\mu$m.
We propose an explanation for the origin of this discrepancy in this 
Letter and discuss the implications for the calibration of the SWS 
database and the impact on calibration of future infrared missions.

\section{Data Format and Analysis \label{sec.data}}

Data from the SWS are publically available in AAR format.  This
format corrects the spectra, as far as possible, for the relative
spectral response of the detectors, differences in gains between
individual detectors and a variety of other issues, as described by
\cite{lee01}.  The files are organized in 12 spectral segments,
ranging from 1 to 4 for each of the 4 different detector bands.
Each spectral segment contains interleaved spectra from the 12 
individual detectors, scanned in the direction of both increasing 
and decreasing wavelength, giving a total of 24 separate spectra 
for each spectral segment.  Thus the processing of the full 
wavelength range from an AAR file requires the combination of 288 
individual spectra into one.

We start with data from the Off-line Processing
(OLP) pipeline 10.0, available in June 2001, which was intended to be 
the final release. This version includes the latest attempts to remove
the memory effect from the Si:As data (4--12~$\mu$m) (Kester et al.
2001).  Although these attempts have significantly improved the data,
residual effects remain, and one can see that the spectra presented
here tend to diverge from the expected result as the data approach
12~$\mu$m.  However, we have high confidence in the quality of the
data at shorter wavelengths, where the flux levels from the sources
are much stronger and where the discrepancies between the SWS data
and the reference spectra occur.

\begin{figure*}
\plotone{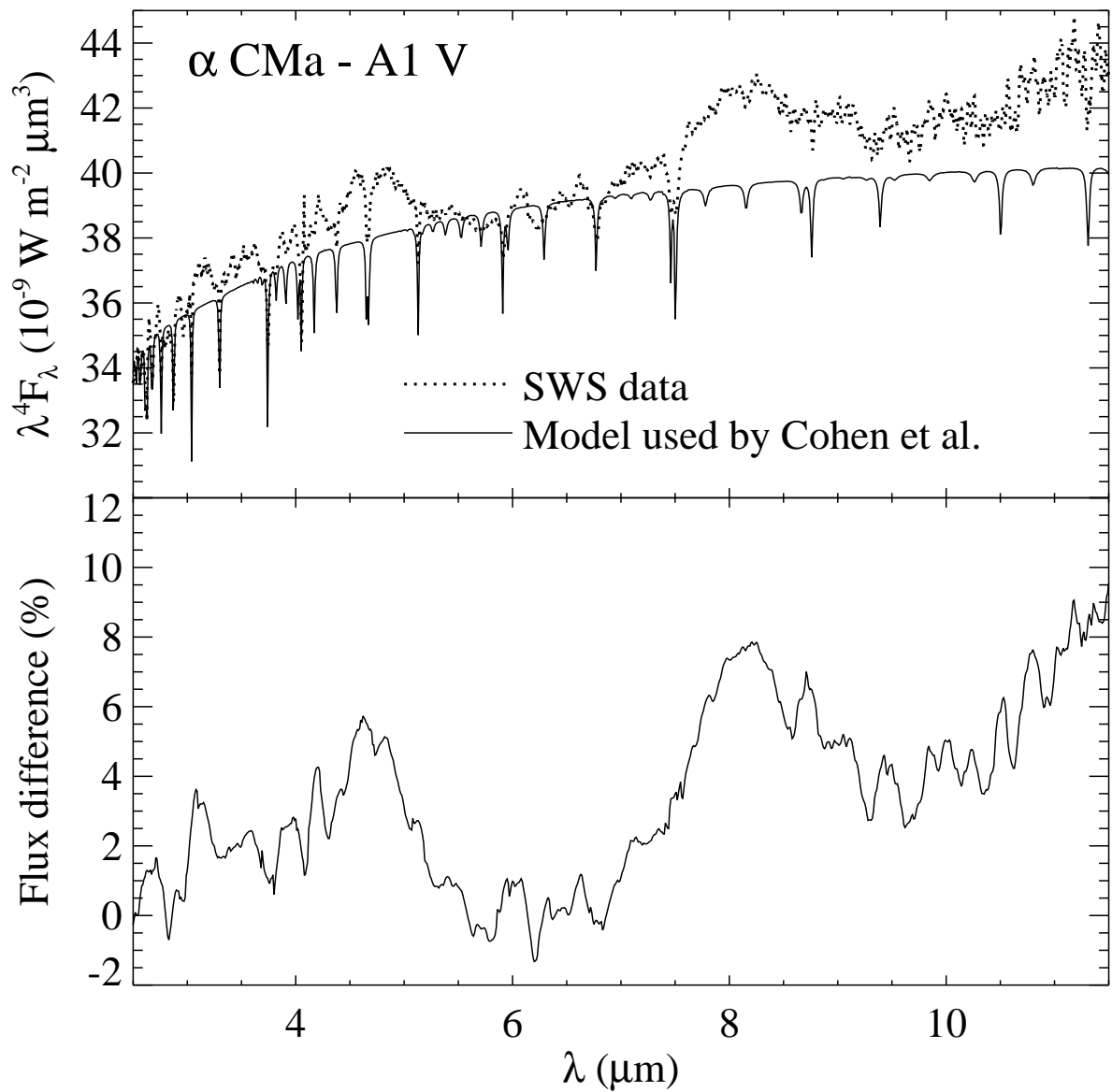}
\caption{A comparison of SWS data for $\alpha$ CMa ({\it dotted
line}) with the Kurucz model used by Cohen et al. 
(1992a) ({\it solid line}).  The spectra have been normalized between 
6 and 7~$\mu$m and are plotted in units of $\lambda^4 F_{\lambda}$, 
which clearly shows deviations from the reference spectrum at the CO 
and SiO bands.  The bottom panel plots the difference between the
two as a percentage of the flux density of the Kurucz model.}
\label{fig.1}
\end{figure*}

Sloan et al. (2002) will present details of the algorithm used to
combine the 288 individual spectra into one.  We have tested for the
possibility that the algorithm might be responsible for the artifacts
seen in Figure \ref{fig.1} by processing the AAR data manually with 
the Infrared Spectral Analysis Package (ISAP) following the standard 
procedures.  The two methods produce similar results, which indicates 
that the problem is intrinsic to the spectrum released by \iso\ and 
not to a specific method of processing.  
The artifact peaking at 4.5~$\mu$m also appears in Figures 6 and 7 of 
\cite{dec01}, although the wavelength stretch makes it more difficult 
to notice.

\section{Discussion \label{sec.disc}}

The divergence between the SWS spectrum of $\alpha$ CMa and the
Kurucz model is characterized by two broad ``emission'' features, at
$\sim$4--5~\mum\ and $\sim$8~\mum.  (The rise at longer wavelengths
is caused by the memory effect mentioned in Sec. \ref{sec.data}.)
The feature at 8~$\mu$m is reminiscent of the 8~$\mu$m emission
artifact found in the data from the Low Resolution Spectrometer (LRS)
aboard \iras\ by Cohen et al. (1992b).  The LRS database was
originally calibrated assuming that $\alpha$ Tau could be
characterized as a $10^4$ K blackbody.  However, since the spectrum
of $\alpha$ Tau (a K5 III) contains significant absorption from the
SiO fundamental band at 8~$\mu$m, this error propagated to the entire
LRS database, producing an apparent emission feature in the spectra
of $\alpha$ CMa, $\alpha$ Lyr, and other early-type stars.

Fifteen stars have served as spectral calibrators for the SWS 
\citep{sch96,lee01,shi01}.  Ten of them are K and M giants, which
show strong molecular absorption bands in their spectra.  The
dominant bands peak at wavelengths of $\sim$2.4~\mum\ (CO overtone),
$\sim$4.1~$\mu$m (SiO overtone), $\sim$4.5~$\mu$m (CO fundamental,
blended with the weaker SiO overtone), and $\sim$8.0~$\mu$m (SiO
fundamental).  For the assumed spectrum of each of these stars in OLP 10.0, 
the SWS instrument team chose to use synthetic models of the stars from 
2 to 12~$\mu$m. For longer wavelengths, a composite spectrum 
from Cohen et al. where available or a template spectrum from 
\cite{cwc99} where no composite existed were used.  The two sets of
data were spliced together at 12~$\mu$m \citep{sch96,lee01,shi01}. 
Previous pipeline releases used the Cohen composites and templates for the
entire wavelength range.

\begin{figure*}
\plotone{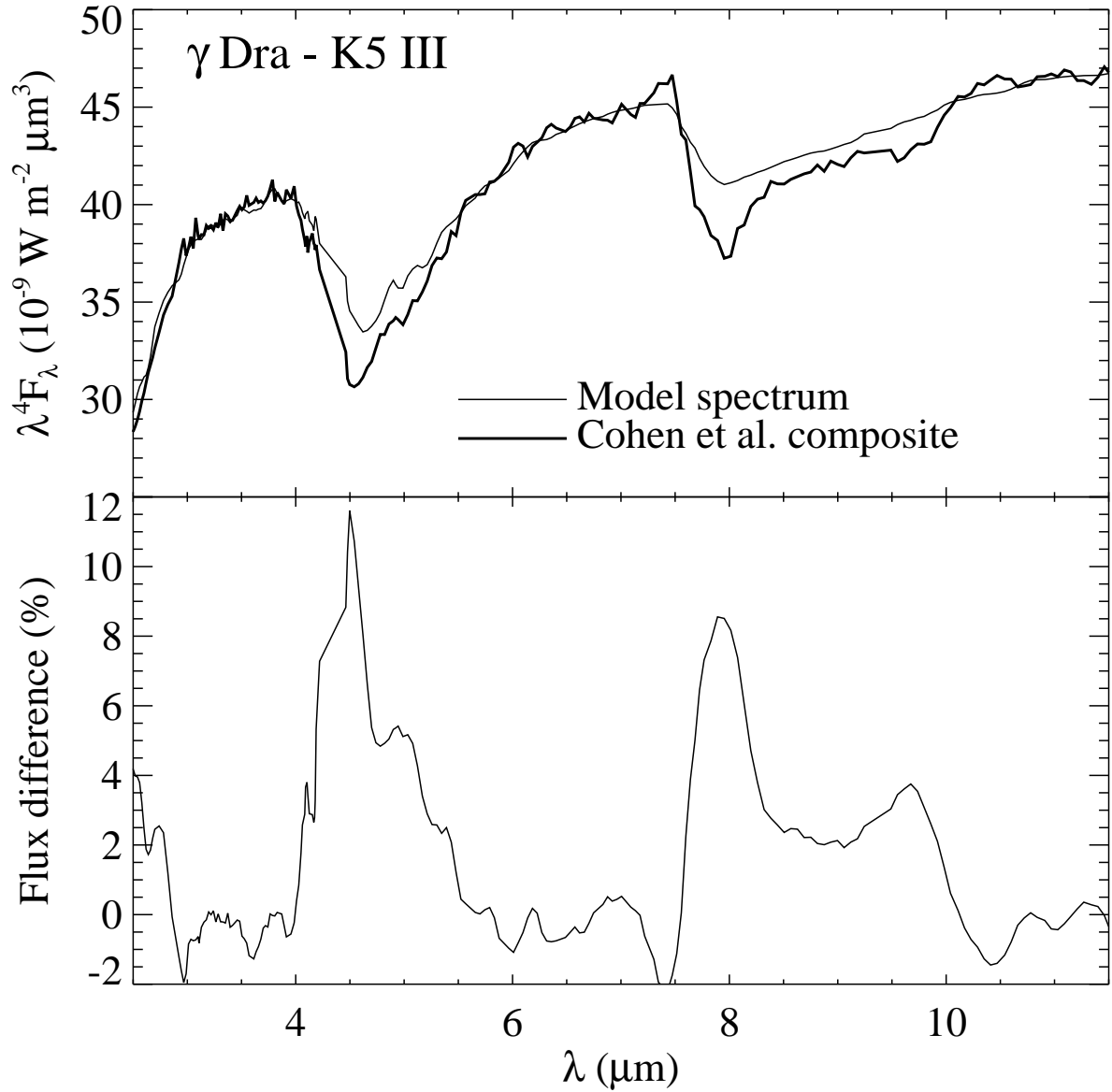}
\caption{A comparison between the model of $\gamma$ Dra used to
calibrate the SWS data ({\it thin line}) and the composite spectrum
calibrated by Cohen et al. (1996b) ({\it thick line}).  The spectra
have been normalized between 6 and 7~$\mu$m and the model resampled to
match the composite.  The data for the model have been adapted from 
Fig. 8 and 9 in \cite{dec01}.}
\label{fig.2}
\end{figure*}

The K5 giant $\gamma$ Dra is a typical example of an SWS calibrator.
Figure \ref{fig.2} compares the synthetic spectrum of this source
\citep{dec00,dec01} to the observationally calibrated composite
spectrum \citep{cwc96b}.  The differences between the two spectra 
(the lower panel) are similar to the artifacts seen in Figure 
\ref{fig.1}.  \cite{dec01} compares the composite spectra of several 
other SWS calibrators to the synthetic spectra (in her Figure 9), 
and most of them show similar deviations\footnote{\cite{dec01} notes the 
discrepancy between the model and the composite spectrum of 
$\gamma$ Dra, but incorrectly attributes it to the substitution of data of 
$\alpha$ Tau for $\gamma$ Dra in this wavelength regime.  
In the header to the file containing the composite spectral data for 
$\gamma$ Dra, \cite{cwc96b} state that they calibrated this spectral 
region using spectral ratios of $\gamma$ Dra to $\alpha$ Boo 
obtained on the KAO.}.  It is highly unlikely 
that an instrumental effect (such as the memory effect) would 
manifest itself precisely at the wavelengths of these photospheric 
absorption bands.

The coincidence of the emission artifacts apparent in Figure 
\ref{fig.1} with the molecular bands in K and M stars and the
similarity of the artifacts to the deviations between the model 
and composite spectrum in Figure \ref{fig.2} lead us to 
hypothesize that the artifacts result from an underestimate of the 
depth of these bands in the synthetic spectra.

\begin{figure*}
\plotone{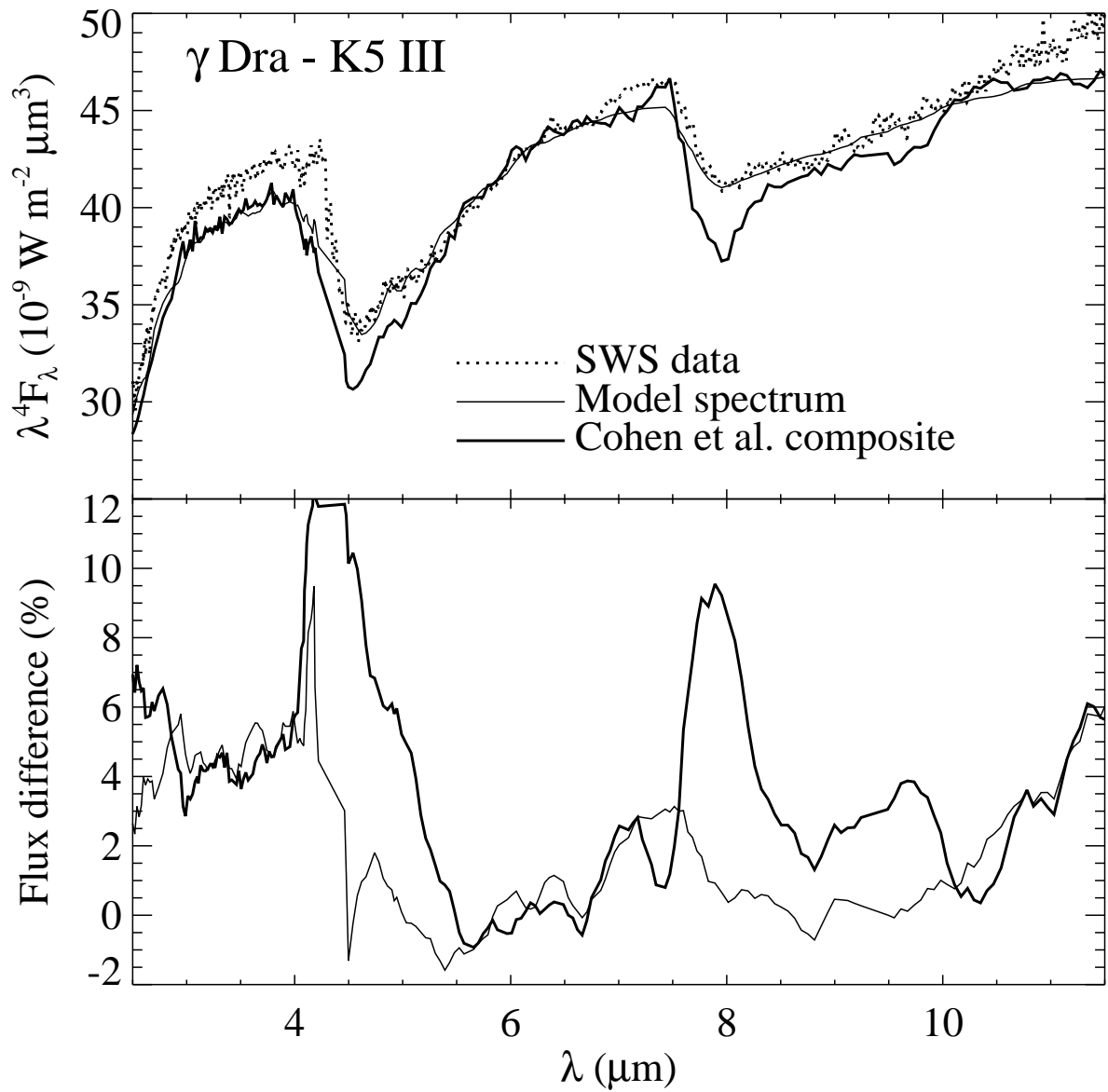}
\caption{Comparing SWS data for $\gamma$ Dra ({\it dotted line}) with 
calibrator spectra.  The upper panel is as in Fig. 2, but with a 
spectrum from the SWS superimposed ({\it dotted line}).  The lower 
panel compares the difference between the SWS data and the model ({\it 
thin line}) to the difference between the SWS data and the composite
from Cohen et al. (1996b) ({\it thick line}).}
\label{fig.3}
\end{figure*}

Figure \ref{fig.3} compares the data from the SWS to both the
model and composite spectrum of $\gamma$ Dra.  Although not perfect, 
{\em the SWS spectrum, as calibrated in OLP 10.0,  matches the 
model of $\gamma$ Dra significantly better than the Cohen composite}.  
Additional spectra of $\gamma$ 
Dra from the SWS database, taken at different times during the 
mission and at different spectral resolutions, as well as spectra
of other cool calibrators, produce similar results.  

\begin{figure*}
\plotone{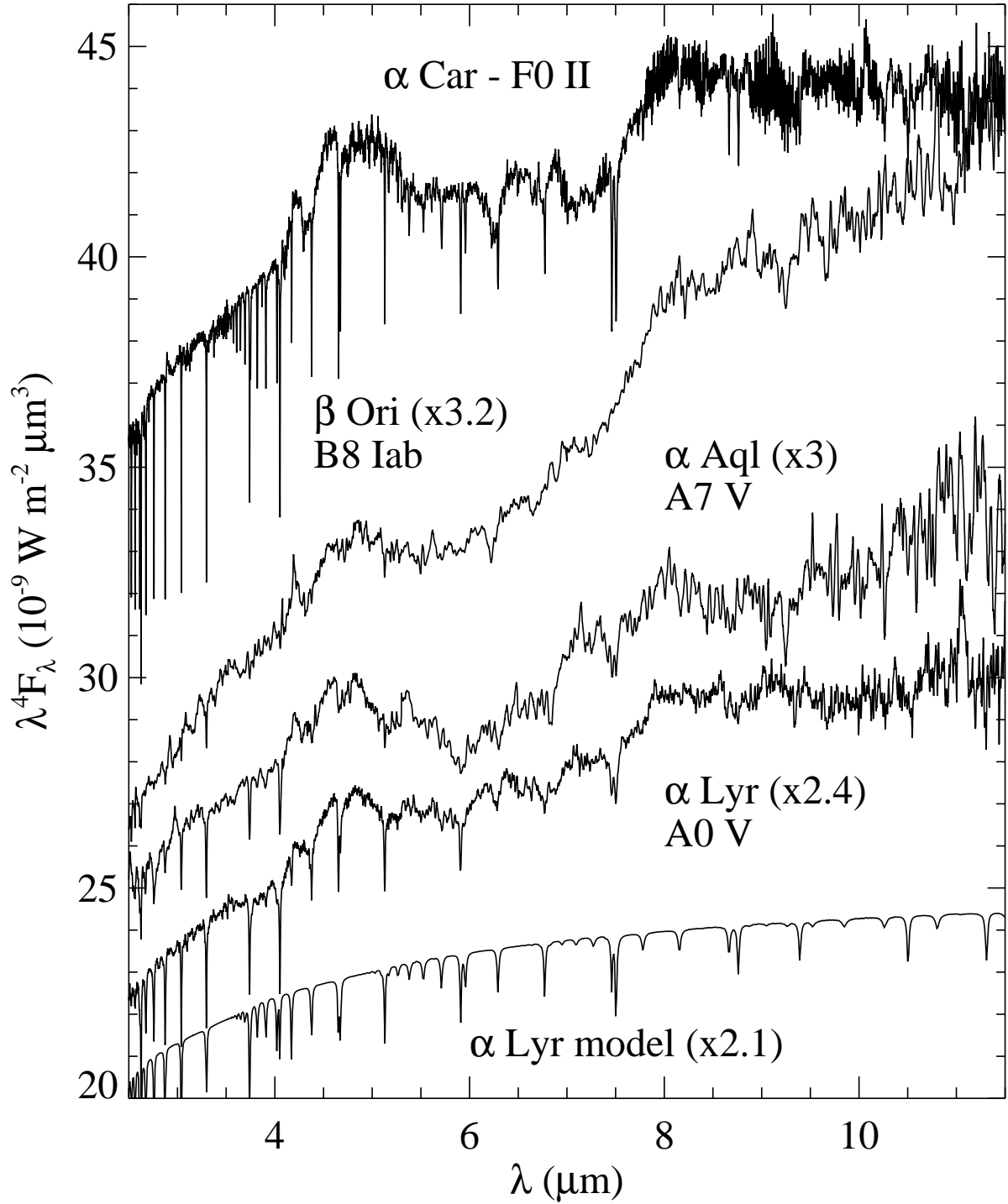}
\caption{Several SWS spectra of bright early-type naked stars.
Most have been multiplied by a constant (given in parentheses).
The bottom spectrum is the Kurucz model of $\alpha$ Lyr used by
\cite{cwb92a}.}
\label{fig.4}
\end{figure*}

If the synthetic spectra used to calibrate the SWS generally 
underestimate the strength of the molecular bands, then this 
miscalibration should propagate to the entire SWS database.  As 
Figure \ref{fig.4} shows, the artifacts at 4.5 and 8~$\mu$m are 
readily seen in bright spectra from early-type stars, including
the reference standard $\alpha$ Lyr, which should show neither 
emission nor absorption from molecular bands.  In the spectrum
of the average giant with no composite spectrum for comparison,
the effect of the miscalibration will be more subtle and more
difficult to recognize, since it will only reduce the apparent
depth of the absorption bands.

In Figure \ref{fig.3} the difference between the SWS data and the 
composite spectrum for $\gamma$ Dra exhibits the largest excess 
between $\sim$4.0 and 4.7~\mum.  While an excess clearly exists at 
these wavelengths in the spectra of the hot stars in Figure
\ref{fig.4}, the confidence level of the magnitude of this excess
shown in Figure \ref{fig.3} is not high.  Because the atmosphere is
almost opaque in the core of the CO$_2$ band at 4.3~\mum, even at
aircraft altitudes,  the uncertainty in the spectral composite is 
typically $\sim$6\% from 4.22 to 4.58 \mum\  and peaks at 10\% at 
4.22 \mum\ \citep{cwc96b}. Roughly half of the composites, including
$\gamma$ Dra, have a data gap at $\sim$4.2-4.4 \mum\ due to the 
atmospheric band.  The spectral artifact is, however, much broader than
the data gap and larger than the uncertainties can account for.
The uncertainties in the composite spectrum  at the other feature are 
$\la$2\%.

\begin{figure*}
\plotone{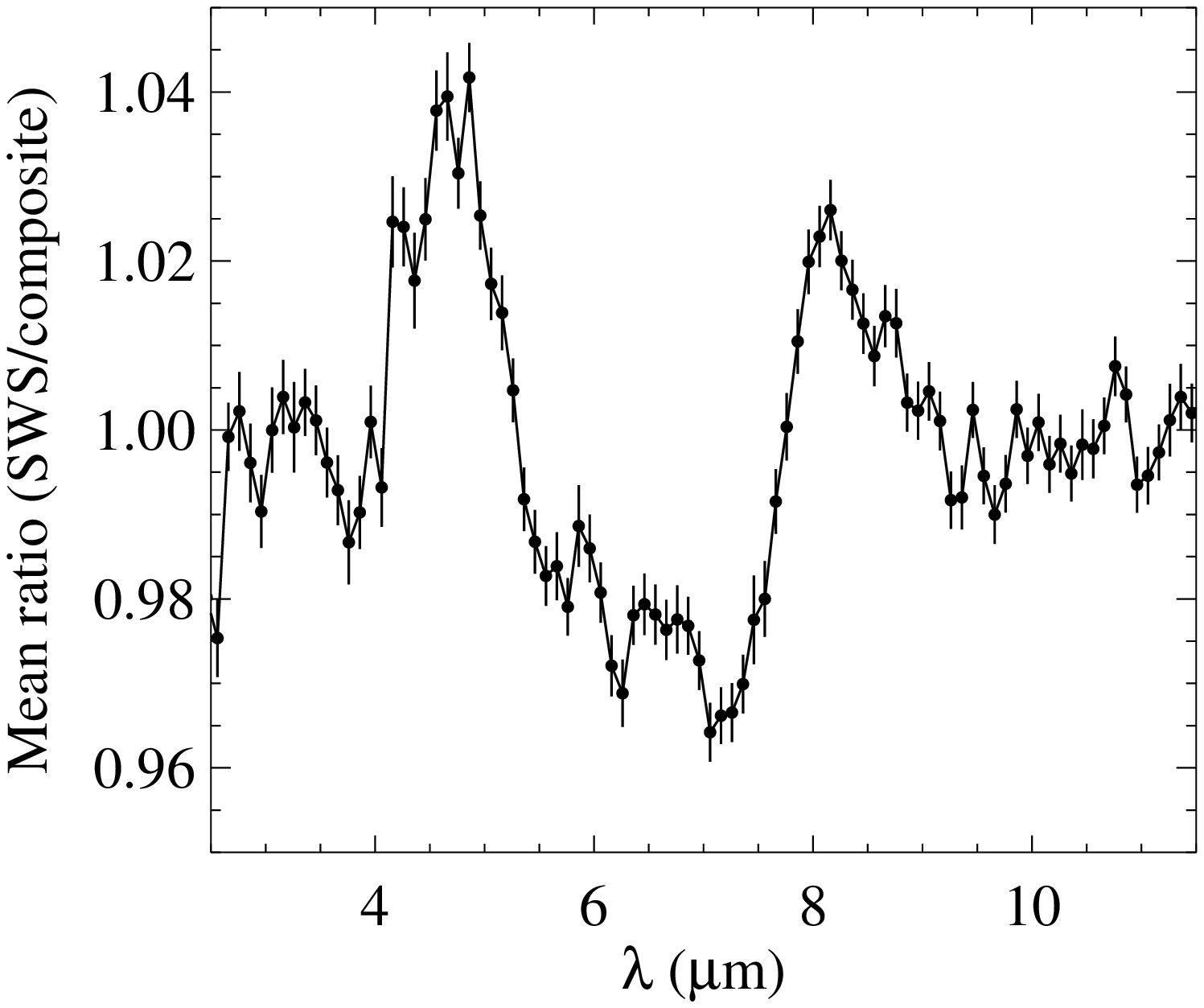}
\caption{The weighted mean of the deviations between the SWS
spectra of the standard stars in common with the composites
and models of Cohen et al.  Both the SWS data and the
composites were resampled to 0.1~\mum\ spacing and placed on 
the same wavelength grid. Error bars are one sigma.}
\label{fig.5}
\end{figure*}

To reduce the noise from a single comparison, we  calculated
the weighted average of the ratios of all of the spectra common
between the sources used to spectroscopically calibrate the SWS
and the composite and reference spectra from Cohen et al.  This
weighted mean, displayed in Figure \ref{fig.5}, provides an
estimate for the modifications needed in the OLP 10.0 data.  It also
provides an assessment of the reliability of the synthetic spectra
relative to the absolutely calibrated composite
spectra.

In light of our findings, the SWS team has revised their calibration
strategy and the \iso\ Data Center has agreed to an additional data
release, OLP 10.1.  By basing the new spectral calibration only on 
stars earlier than K, the worst of the discrepancy between the 
synthetic spectra and the Cohen spectra should be mitigated. The 
Decin models agree with the Cohen composites, templates, or Kurucz 
models (as appropriate) for the five stars (two As, one F, and two Gs) 
which meet this criterion (R. Shipman 2001, private communication).

\section{Conclusions}

We have identified spectral artifacts in the vicinity of the CO
and SiO bands of late-type standard stars that appear in all OLP 10.0
spectra.  These artifacts constitute a known systematic bias over
a fairly large spectral range, and removing them is relatively
straightforward. While OLP 10.0 was to have been
the final version, an additional release, OLP 10.1, has been made
to correct the database based on our findings.

The SWS calibrations between 2 and 12~$\mu$m were predominantly based on 
stellar atmospheric models of cool stars.  These models have been used 
with great success in analysis of physical properties of cool stars 
\citep[e.g.][]{dwe00}.  Model spectra of A stars, which contain only 
atomic lines and no molecular bands, appear to be well founded. Synthetic
spectra can also achieve higher spectral resolution than that afforded by
the composites. However, 
the artifacts in the SWS database that mirror the dominant molecular 
absorption features in the 4--10~$\mu$m range indicate that synthetic 
spectra of cool giants require further progress before they can be 
used for definitive spectral calibration. 

The Infrared Spectrograph (IRS) on the upcoming Space Infrared
Telescope Facility (SIRTF) faces issues similar to those encountered by
the SWS team, with the added complication that all of the commonly
used spectral standards will be far too bright for use on the IRS.
Until the commission of the Stratospheric Observatory for Infrared
Astronomy (SOFIA), it will prove difficult to observationally
calibrate standards faint enough for use by the IRS.  Until this
happens, the IRS will need to rely on models for spectral calibration.
If models of later stars are required, improvements in the production 
of synthetic spectra will also be necessary.

\acknowledgements
We wish to thank Russ Shipman for many useful discussions regarding 
the calibration process.  Thijs de Graauw provided support for this
work through the USAF/SRON MOU.   One author (KEK) was supported by 
the National Research Council via a Research Associateship through 
the Air Force Office of Scientific Research.  This research has made 
use of NASA's Astrophysics Data System Abstract Service.

\end{document}